\documentclass[prc,preprint]{revtex4}
\usepackage{graphics}

\newcommand{\be}{\begin{equation}}
\newcommand{\ee}{\end{equation}}

\begin{document}
\title{{\bf Lowest order constrained variational calculation of
polarized neutron matter at finite temperature}}

\author{{\bf G.H. Bordbar $^{1,3}$ \footnote{Corresponding author. E-mail:
bordbar@physics.susc.ac.ir}} and {\bf M.
Bigdeli$^{2,3}$\footnote{E-mail: m\underline \
bigdeli@znu.ac.ir}}}
 \affiliation{ $^1$Department of Physics,
Shiraz University,
Shiraz 71454, Iran\footnote{Permanent address},\\
 $^2$Department of Physics, Zanjan University, P.O. Box 45195-313, Zanjan,
 Iran\\  and\\
$^3$Research Institute for Astronomy and Astrophysics of Maragha,\\
P.O. Box 55134-441, Maragha, Iran}



\begin{abstract}
Some properties of the polarized neutron matter at finite
temperature has been studied using the lowest  order constrained
variational (LOCV) method with the $AV_{18}$ potential. Our
results indicate that spontaneous transition to the ferromagnetic
phase does not occur. Effective mass, free energy, magnetic
susceptibility, entropy and the equation of state of the polarized
neutron matter at finite temperature are also calculated. A
comparison is also made between our results and those of other
many-body techniques.
\end{abstract}

\maketitle
\section{INTRODUCTION}
The spontaneous phase transition to a ferromagnetic state in the
neutron matter is of particular interest in astrophysics.
Specially, this transition could have important consequences for
the physical origin of magnetic field of pulsars, that are
believed to be rapidly rotating neutron stars with strong surface
magnetic fields in the range of $10^{12} -10^{13}$ Gauss
\cite{shap,paci,gold,navarro}, and also the evolution of a
protoneutron star. A hot neutron star is born within a short time
after supernovae explosion. In this stage (protoneutron star), the
interior temperature of a neutron star is of the order 20-50 MeV
\cite{burro}. Therefore, the study of magnetic properties of
polarized neutron matter at finite temperature is of special
interest in the description of protoneutron stars.

There exist several possibilities of the generation of the
magnetic field in a neutron star. From the nuclear physics point
of view, one is the possible existence of a phase transition to a
ferromagnetic state at densities corresponding to the
theoretically stable neutron stars and, therefore, of a
ferromagnetic core in the liquid interior of such compact objects.
Such a possibility has been studied by several authors using
different theoretical approaches [6-29], but the results are still
contradictory. In most calculations, neutron star matter is
approximated by pure neutron matter, as proposed just after the
discovery of pulsars. Whereas some calculations, like those of
based on hard sphere gas model \cite{brown,rice}, Skyrmelike
interactions \cite{vida},  Reid soft-core potential \cite{pandh}
and relativistic Dirac-Hartree-Fock approximation with an
effective nucleon-meson Lagrangian \cite{marcos} showed that
neutron matter becomes ferromagnetic at some densities.
Others, like recent Monte Carlo \cite{fanto} and
Brueckner-Hartree-Fock calculations [21-23] using modern two-body
and three-body realistic interactions show no indication of
ferromagnetic transition at any density for neutron matter and
asymmetrical nuclear matter.

Most of the studies of the ferromagnetic transition in neutron
matter and nuclear matter have been done at zero temperature. The
properties of polarized neutron matter both at zero and finite
temperature have been  studied by several authors
\cite{apv,dapv,bprrv}. Bombaci et al. (BPRRV) \cite{bprrv} have
studied the properties of polarized neutron matter within the
framework of the Brueckner-Hartree-Fock formalism using the
$AV_{18}$ nucleon-nucleon interaction. Their results show no
indication of a ferromagnetic transition at any density and
temperature. Lopez-Val et al. \cite{dapv} have used the D1 and D1P
parameterization of the Gogny interaction and the results of their
calculation show two different behaviors: whereas the D1P force
exhibits a ferromagnetic transition at density of $\rho\sim1.31
fm^{-3}$ whose onset increases with temperature, no sign of such a
transition is found for D1 at any density and temperature. Rios
and Polls \cite{apv} have used Skyrme-like interactions and their
results indicate the occurrence of a ferromagnetic phase of
neutron matter.

Recently, we have computed the properties of polarized neutron
matter \cite{bordbig}, polarized symmetrical nuclear matter
\cite{bordbig2}, such as total energy, magnetic susceptibility,
pressure, etc at zero temperature using the microscopic
calculations employing the lowest  order constrained variational
(LOCV) method with the $AV_{18}$ potential. We have also
calculated the properties of spin polarized asymmetrical nuclear
matter and neutron star matter \cite{bordbig3} using the LOCV
method employing the $AV_{18}$ \cite{wiring}, $Reid93$ \cite{R93},
$UV_{14}$ \cite{UV14} and $AV_{14}$ \cite{AV14} potentials. We
have concluded that the spontaneous phase transition to a
ferromagnetic state in the neutron matter, symmetrical and
asymmetrical nuclear matter and neutron star matter does not
occur.

In the present work, we study the properties of polarized neutron
matter at finite temperature using the LOCV technique employing
the $AV_{18}$ potential.


\section{LOCV FORMALISM FOR POLARIZED HOT NEUTRON MATTER}
We consider a system of $N$ interacting neutrons with $N_1$ spin
up and $N_2$ spin down neutrons. The total number density ($\rho$)
and spin asymmetry parameter ($\delta$) are defined as
\begin{eqnarray}
\rho&=&\rho_1+\rho_2,\nonumber\\
\delta&=&\frac{N_1-N_2}{N}.
\end{eqnarray}
$\delta$ shows the spin ordering of matter which can have a value
in the range of $\delta=0.0$ (unpolarized matter) to $\delta=1.0$
(fully polarized matter).

Now, we consider a trial many-body wave function of the form
\begin{eqnarray}
     \psi=F\phi,
 \end{eqnarray}
where $\phi$ is the uncorrelated ground state wave function
(simply the Slater determinant of plane waves) of $N$ independent
neutrons and $F=F(1\cdots N)$ is an appropriate N-body correlation
operator which can be replaced by a Jastrow form i.e.,
\begin{eqnarray}
    F=\textsf{S}\prod _{i>j}f(ij),
 \end{eqnarray}
in which {\textsf{S}} is a symmetrizing operator. We consider a
cluster expansion of the energy functional up to the two-body
term,
 \begin{eqnarray}\label{tener}
           E([f])=\frac{1}{N}\frac{\langle\psi|H\psi\rangle}
           {\langle\psi|\psi\rangle}=E _{1}+E _{2}\cdot
 \end{eqnarray}
For hot neutron matter, the one-body term $E _{1}$ is
\begin{eqnarray}\label{ener1}
               E _{1}=\sum _{i=1,2}\varepsilon_{i}\cdot
 \end{eqnarray}
Labels 1 and 2 are used instead of spin up and spin down neutrons,
respectively, and $\varepsilon_{i}$ is
 \begin{eqnarray}
               \varepsilon_{i}=\sum _{k}
               \frac{\hbar^{2}{k^2}}{2m}n_{i}(k,T,\rho_{i}),
                \end{eqnarray}
where $n_{i}(k,T,\rho_{i})$ is the Fermi-Dirac distribution
function,
\begin{eqnarray}
              n_{i}(k,T,\rho_{i})=\frac{1}{e^{\beta[\epsilon_{i}
              (k,T,\rho_{i})-\mu_{i}(T,\rho_{i})]}+1}\cdot
                \end{eqnarray}
In the above equation $\beta=\frac{1}{T}$ and $\mu_{i}$ being the
chemical potential, $\rho_{i}$ is the number density and
$\epsilon_{i}$ is the single particle energy of a neutron with
spin projection $i$.

In our formalism, the single particle energy, $\epsilon_{i}$, of a
neutron with momentum $k$ and spin projection $i$ is approximately
written in terms of effective mass as  \cite{apv,dapv}
 \begin{eqnarray}
               \epsilon_{i}(k,T,\rho_{i})=
               \frac{\hbar^{2}{k^2}}{2m^{*}_{i}(\rho,T)}+U_i(T,\rho_i)\cdot
                \end{eqnarray}
In fact, we use a quadratic approximation for single particle
potential, incorporated in the single particle energy as a
momentum independent effective mass. $U_i(T,\rho_i)$ is the
momentum independent single particle potential. The effective
mass, $m^{*}_{i}$, is determined variationally
\cite{mod,mod95,mod97,modbord,fp}.

The chemical potentials, $\mu_{i}$, at any adopted values of the
temperature ($T$), number density ($\rho_i$) and spin polarization
($\delta$), are determined by applying the constraint,
 \begin{eqnarray}\label{chpt}
              \sum _{k}
               n_{i}(k,T,\rho_{i})=N_{i}\cdot
                \end{eqnarray}
This is an implicit equation which can be solved numerically.

 The two-body energy $E_{2}$ is
\begin{eqnarray}
    E_{2}&=&\frac{1}{2A}\sum_{ij} \langle ij\left| \nu(12)\right|
    ij-ji\rangle,
 \end{eqnarray}
where \\ $\nu(12)=-\frac{\hbar^{2}}{2m}[f(12),[\nabla
_{12}^{2},f(12)]]+f(12)V(12)f(12)$, $f(12)$ and $V(12)$ are the
two-body correlation and potential. For the two-body correlation
function, $f(12)$, we consider the following form
\cite{borda,bordb}:
\begin{eqnarray}
f(12)&=&\sum^3_{k=1}f^{(k)}(12)O^{(k)}(12),
\end{eqnarray}
where, the operators $O^{(k)}(12)$ are given by
\begin{eqnarray}
O^{(k=1-3)}(12)&=&1,\ (\frac{2}{3}+\frac{1}{6}S_{12}),\
(\frac{1}{3}-\frac{1}{6}S_{12}),
\end{eqnarray}
and $S_{12}$ is the tensor operator. After doing some algebra, we
find the following equation for the two-body energy:
\begin{eqnarray}\label{ener2}
    E_{2} &=& \frac{2}{\pi ^{4}\rho }\left( \frac{h^{2}}{2m}\right)
    \sum_{JLSS_{z}}\frac{(2J+1)}{2(2S+1)}[1-(-1)^{L+S+1}]\left| \left\langle
\frac{1}{2}\sigma _{z1}\frac{1}{2}\sigma _{z2}\mid
SS_{z}\right\rangle \right| ^{2} \int dr\left\{\left [{f_{\alpha
}^{(1)^{^{\prime }}}}^{2}{a_{\alpha
}^{(1)}}^{2}(k_{f}r)\right.\right. \nonumber
\\&& \left.\left. +\frac{2m}{h^{2}}(\{V_{c}-3V_{\sigma } +V_{\tau }-3V_{\sigma
\tau }+2(V_{T}-3V_{\sigma \tau }) +2V_{\tau z}\}{a_{\alpha
}^{(1)}}^{2}(k_{f}r)\right.\right. \nonumber \\&&\left.\left.
+[V_{l2}-3V_{l2\sigma } +V_{l2\tau }-3V_{l2\sigma \tau
}]{c_{\alpha }^{(1)}}^{2}(k_{f}r))(f_{\alpha }^{(1)})^{2}\right ]
+\sum_{k=2,3}\left[ {f_{\alpha }^{(k)^{^{\prime }}}}^{2}{a_{\alpha
}^{(k)}}^{2}(k_{f}r)\right.\right. \nonumber \\&&\left. \left.
+\frac{2m}{h^{2}}( \{V_{c}+V_{\sigma }+V_{\tau } +V_{\sigma \tau
}+(-6k+14)(V_{tz}+V_{t})-(k-1)(V_{ls\tau }+V_{ls})\right.\right.
\nonumber
\\&&\left.\left. +[V_{T}+V_{\sigma \tau }+(-6k+14)V_{tT}] [2+2V_{\tau
z}]\}{a_{\alpha }^{(k)}}^{2}(k_{f}r)\right.\right. \nonumber
\\&&\left.\left. +[V_{l2}+V_{l2\sigma } +V_{l2\tau }+V_{l2\sigma \tau
}]{c_{\alpha }^{(k)}}^{2}(k_{f}r)+[V_{ls2}+V_{ls2\tau }]
{d_{\alpha }^{(k)}}^{2}(k_{f}r)) {f_{\alpha }^{(k)}}^{2}\right
]\right. \nonumber \\&&\left. +\frac{2m}{h^{2}}\{V_{ls}+V_{ls\tau
}-2(V_{l2}+V_{l2\sigma }+V_{l2\sigma \tau } +V_{l2\tau
})-3(V_{ls2} +V_{ls2\tau })\}b_{\alpha }^{2}(k_{f}r)f_{\alpha
}^{(2)}f_{\alpha }^{(3)}\right. \nonumber \\&&\left.
+\frac{1}{r^{2}}(f_{\alpha }^{(2)} -f_{\alpha
}^{(3)})^{2}b_{\alpha }^{2}(k_{f}r)\right\},
 \end{eqnarray}
where $\alpha=\{J,L,S,S_z\}$ and the coefficient  ${a_{\alpha
}^{(1)}}^{2}$, etc., are defined as
\begin{eqnarray}\label{a1}
     {a_{\alpha }^{(1)}}^{2}(r,\rho,T)=r^{2}I_{L,S_{z}}(r,\rho,T),
 \end{eqnarray}
\begin{eqnarray}
     {a_{\alpha }^{(2)}}^{2}(r,\rho,T)=r^{2}[\beta I_{J-1,S_{z}}(r,\rho,T)
     +\gamma I_{J+1,S_{z}}(r,\rho,T)],
 \end{eqnarray}
\begin{eqnarray}
           {a_{\alpha }^{(3)}}^{2}(r,\rho,T)=r^{2}[\gamma I_{J-1,S_{z}}(r,\rho,T)
           +\beta I_{J+1,S_{z}}(r,\rho,T)],
      \end{eqnarray}
\begin{eqnarray}
     b_{\alpha }^{(2)}(r,\rho,T)=r^{2}[\beta _{23}I_{J-1,S_{z}}(r,\rho,T)
     -\beta _{23}I_{J+1,S_{z}}(r,\rho,T)],
 \end{eqnarray}
\begin{eqnarray}
         {c_{\alpha }^{(1)}}^{2}(r,\rho,T)=r^{2}\nu _{1}I_{L,S_{z}}(r,\rho,T),
      \end{eqnarray}
\begin{eqnarray}
        {c_{\alpha }^{(2)}}^{2}(r,\rho,T)=r^{2}[\eta _{2}I_{J-1,S_{z}}(r,\rho,T)
        +\nu _{2}I_{J+1,S_{z}}(r,\rho,T)],
 \end{eqnarray}
\begin{eqnarray}
       {c_{\alpha }^{(3)}}^{2}(r,\rho,T)=r^{2}[\eta _{3}I_{J-1,S_{z}}(r,\rho,T)
       +\nu _{3}I_{J+1,S_{z}}(r,\rho,T)],
 \end{eqnarray}
\begin{eqnarray}
     {d_{\alpha }^{(2)}}^{2}(r,\rho,T)=r^{2}[\xi _{2}I_{J-1,S_{z}}(r,\rho,T)
     +\lambda _{2}I_{J+1,S_{z}}(r,\rho,T)],
 \end{eqnarray}
\begin{eqnarray}\label{d2}
     {d_{\alpha }^{(3)}}^{2}(r,\rho,T)=r^{2}[\xi _{3}I_{J-1,S_{z}}(r,\rho,T)
     +\lambda _{3}I_{J+1,S_{z}}(r,\rho,T)],
 \end{eqnarray}
with
\begin{eqnarray}
          \beta =\frac{J+1}{2J+1},\ \gamma =\frac{J}{2J+1},\
          \beta _{23}=\frac{2J(J+1)}{2J+1},
 \end{eqnarray}
\begin{eqnarray}
       \nu _{1}=L(L+1),\ \nu _{2}=\frac{J^{2}(J+1)}{2J+1},\
       \nu _{3}=\frac{J^{3}+2J^{2}+3J+2}{2J+1},
      \end{eqnarray}
\begin{eqnarray}
     \eta _{2}=\frac{J(J^{2}+2J+1)}{2J+1},\ \eta _{3}=
     \frac{J(J^{2}+J+2)}{2J+1},
 \end{eqnarray}
\begin{eqnarray}
     \xi _{2}=\frac{J^{3}+2J^{2}+2J+1}{2J+1},\
     \xi _{3}=\frac{J(J^{2}+J+4)}{2J+1},
 \end{eqnarray}
\begin{eqnarray}
     \lambda _{2}=\frac{J(J^{2}+J+1)}{2J+1},\
     \lambda _{3}=\frac{J^{3}+2J^{2}+5J+4}{2J+1},
 \end{eqnarray}
and
\begin{eqnarray}
       I_{J,S_{z}}(r,\rho,T)=\frac{1}{2\pi^{6}\rho^2}\int dk_{1}dk_{2}n_{i}
       (k_{1},T,\rho_{i})n_{j}(k_{2},T,\rho_{j})J_{J}^{2}(|k_{2}-k_{1}|r)\cdot
 \end{eqnarray}
In the above equation $J_{J}(x)$ is the Bessel's function .

Now, we  minimize the two-body energy Eq.(\ref{ener2}), with
respect to the variations in the functions ${f_{\alpha}}^{(i)}$,
but subject to the normalization constraint \cite{bordb},
\begin{eqnarray}
        \frac{1}{A}\sum_{ij}\langle ij\left| h_{S_{z}}^{2}
        -f^{2}(12)\right| ij\rangle _{a}=0\cdot
 \end{eqnarray}
In the case of spin polarized neutron matter, the function
$h_{S_{z}}(r)$ is defined as
\begin{eqnarray}
       h_{S_{z}}(r)&=&\left[ 1-\left( \frac{\gamma_{i}(r)
       }{\rho}\right) ^{2}\right] ^{-1/2};\  S_{z}=\pm1
       \nonumber\\
       &=& 1\ \ \ \ \ \ \ \ \ \ \ \ \ \ \ \ \ \ \ \ \ \ \ \ \ \ \
       \ \ \ ;\ S_{z}= 0
 \end{eqnarray}
 where
\begin{eqnarray}
       \gamma_{i}(r)=\frac{1}{2\pi^{2}}\int n_{i}(k,T,\rho_{i})J_{0}(kr)k^2 dk \cdot
 \end{eqnarray}
From the minimization of the two-body cluster energy, we get a set
of coupled and uncoupled differential equations which are the same
as presented in Ref. \cite{bordb}, with the coefficients replaced
by those indicated in equations (\ref{a1})-(\ref{d2}). We can
obtain correlation functions by solving the differential equations
and then the two-body energy is computed.

 Finally, we must calculate the total free energy per particle,
$F$, to get the macroscopic properties of hot neutron matter,
\begin{eqnarray}\label{free}
       F=E-TS,
 \end{eqnarray}
where $S$ is the entropy per neutron,
 \begin{eqnarray}
              S(\rho,T)=-\frac{1}{N}\sum _{i=1,2}\sum _{k}
               \{[1-n_{i}(k,T,\rho_{i})]\textrm{ln}[1-n_{i}(k,T,\rho_{i})]
               +n_{i}(k,T,\rho_{i}) \textrm{ln} n_{i}(k,T,\rho_{i})\}.
                \end{eqnarray}

We introduce the effective masses, $m^{*}_{i}$, as variational
parameters \cite{mod,mod95,mod97,modbord,fp}.
 In fact, we minimize the free energy with respect to the variations in
 the effective masses and then, we obtain
 the chemical potentials and the effective masses of spin-up and spin-down
 neutrons
 at the minimum point of the free energy.
 This minimization is done numerically.

\section{RESULTS}\label{NLmatchingFFtex}
In Fig. 1, we have plotted the effective mass of spin-up and
spin-down neutrons versus spin polarization($\delta$) for fixed
density $\rho=0.16 fm^{-3}$ at $T=10$ and $T=20$ MeV. We see that
the effective masses of the above components have the same values
at $\delta=0$ and get different values by increasing the spin
polarization. It can be seen that the effective mass of spin-up
neutrons is larger than spin-down neutrons for $\delta > 0$ and
the effective mass of spin-down (spin-up) neutrons decreases
(increases) by increasing the polarization. In Fig. 1, we have
also included the results of BPRRV calculations \cite{bprrv} for
comparison. It is seen that the behavior of the effective mass
versus $\delta$ obtained from our method shows the same properties
as BPRRV calculations \cite{bprrv}.

Our results indicate that the effective masses of spin-up and
spin-down neutrons increase by increasing the temperature.
Whereas, the results of BPRRV \cite{bprrv} and Isayev
\cite{Isayev} show that the effective masses of nucleons in the
spin polarized neutron matter and nuclear matter are decreasing
functions of the temperature.
However, in the references \cite{bprrv} and \cite{Isayev}, the
authors determine the momentum independent effective masses
through the derivatives of the self-consistently calculated single
particle potentials at the corresponding Fermi momenta while, in
our calculations, the effective masses are introduced in the
single particle energies and considered as variational parameters
for the free energy.
We note that increasing the effective mass by increasing the
temperature can be also seen in the results of others where the
procedures for finding the effective mass are the same as ours
\cite{mod,mod95,mod97,fp}.

The behavior of the free energy per particle of the polarized hot
neutron matter versus total number density ($\rho$) for different
values of the spin polarization ($\delta$) at $T=10$ and $T=20$
MeV is shown in Fig. 2. This figure shows that the free energy
increases with increasing both density and polarization. It is
seen that the minimum value of free energy occurs for unpolarized
case ($\delta =0.0$) at any density and at relevant considered
temperature. By comparison, in the two panels of Fig. 2, we see
that the free energy decreases by increasing temperature. It is
also seen that there is no crossing of the free energy curves for
different polarizations. Conversely, by increasing the density,
the difference between the free energy of neutron matter at
different polarizations becomes more appreciable. According to
this result, the spontaneous phase transition to a ferromagnetic
state in the hot neutron matter does not occur. If such a
transition existed, a crossing of the energies of different
polarizations would have been observed at some density, indicating
that the ground state of the system would be ferromagnetic from
that density on.

In Fig. 3, we have presented the free energy per particle as a
function of the quadratic spin polarization ($\delta^2$) for fixed
density $\rho=0.36 fm^{-3}$ at $T=10$ and $T=20$ MeV. It can be
seen that the free energy per particle increases by increasing the
polarization. We see that the variation of the free energy of hot
neutron matter versus $\delta^{2}$ is nearly linear. This
indicates that the ground state of hot neutron matter is
paramagnetic. In Fig. 3, we have compared our results with those
of BPRRV calculations \cite{bprrv}. It is seen that there is an
overall agreement between our results and the results of BPRRV
\cite{bprrv}.

The magnetic susceptibility, $\chi$, which characterizes the
response of a system to the magnetic field is given by
\begin{eqnarray}\label{susep}
   \chi =\frac{\mu ^{2}\rho
}{\left( \frac{\partial^{2}F}{\partial \delta ^{2}}\right)
_{\delta =0}},
\end{eqnarray}
where $\mu$ is the magnetic moment of neutron. In Fig. 4, we have
shown the ratio ${{\chi_{F}}/{\chi}}$ as a function of the total
number density at two values of temperature $T=10$ and $T=20$ MeV.
$\chi_{F}$ is the magnetic susceptibility for a free Fermi gas. As
can be seen from Fig. 4, even at high densities, the ratio
${{\chi_{F}}/{\chi}}$ increases monotonically and continuously as
the density increases for any temperature. This shows that there
is no magnetic instability in hot neutron matter. In Fig. 4, the
results of BPRRV calculations \cite{bprrv} at $T=20$ MeV are also
presented for comparison. We see that our results and BPRRV
results \cite{bprrv} have an agreement at low densities.

In Fig. 5, the difference of the entropy per particle of fully
polarized and unpolarized cases is plotted as a function of the
total number density at $T=10$ and $T=20$ MeV. Fig. 5 shows that
for all relevant densities, this difference has negative values.
According to this result, we can conclude that the fully polarized
case is more ordered than the unpolarized case. In Fig. 5, the
results of BPRRV calculations \cite{bprrv} are also given for
comparison. There is an agreement between our results and the
BPRRV \cite{bprrv} results at low densities.

In Fig. 6, we have plotted the entropy per particle of hot neutron
matter versus spin polarization for fixed density $\rho=0.32
fm^{-3}$ and temperature $T=20MeV$. It is seen that the entropy
decreases as polarization increases with its highest value
occurring for the unpolarized case. To prevent anomalous behavior
of entropy as a function of spin polarization, for a given density
and temperature, a condition for effective masses can be derived
\cite{apv},
\begin{eqnarray}\label{emcon}
  \frac{m^{*}(\rho,\delta=1.0)}{m^{*}(\rho,\delta=0.0)}<2^{2/3},
\end{eqnarray}
where $m^{*}(\rho,\delta=1.0)$ and $m^{*}(\rho,\delta=0.0)$ are
the effective masses of the fully polarized and unpolarized
neutron matter, respectively. From Fig. 1, we have found out that
for density $\rho=0.16 fm^{-3}$, the above ratio at $T=10$ MeV is
$\frac{m_{1}^{*}(\delta=1.0)}{m_{1,2}^{*}(\delta=0.0)}=1.22$ and
at $T=20 MeV$ is
$\frac{m_{1}^{*}(\delta=1.0)}{m_{1,2}^{*}(\delta=0.0)}=1.18$ which
are smaller than the indicated limit. This condition is satisfied
for all other densities explored in this work. Therefor, we can
see that the entropy of polarized case is always  smaller than the
entropy of unpolarized case.

The equation of state of hot polarized neutron matter,
$P(\rho,T,\delta)$, can be simply obtained using
\begin{eqnarray}
      P(\rho,T,\delta)= \rho^{2} \frac{\partial F(\rho,T,\delta)}
      {\partial \rho}
 \end{eqnarray}
In Fig. 7, we have presented the pressure of neutron matter as a
function of the total number density ($\rho$) for different
polarizations at $T=10$ and $T=20$ MeV. We see that the equation
of state becomes stiffer by increasing the polarization.

\section{Summary and Conclusions}
Some thermodynamic properties of the polarized neutron matter at
finite temperature were reexamined using the lowest order
constrained variational (LOCV) method employing the $AV_{18}$
nucleon-nucleon potential. Our main goal was to check the
occurrence of the spontaneous transition to the ferromagnetic
state. We found no indication for the occurrence the ferromagnetic
phase, in agreement with the results of others who used the
different many-body techniques. Effective mass, free energy per
particle, magnetic susceptibility, entropy per particle, and the
equation of state for the polarized neutron matter at finite
temperature were calculated and the effect of polarization on
these properties were examined.

\acknowledgements{This work has been supported by Research
Institute for Astronomy and Astrophysics of Maragha. We wish to
thank Shiraz University and Zanjan University Research Councils.
We also wish to thank A. Poostforush for various useful
discussion.}


\newpage
\begin{figure}

\includegraphics{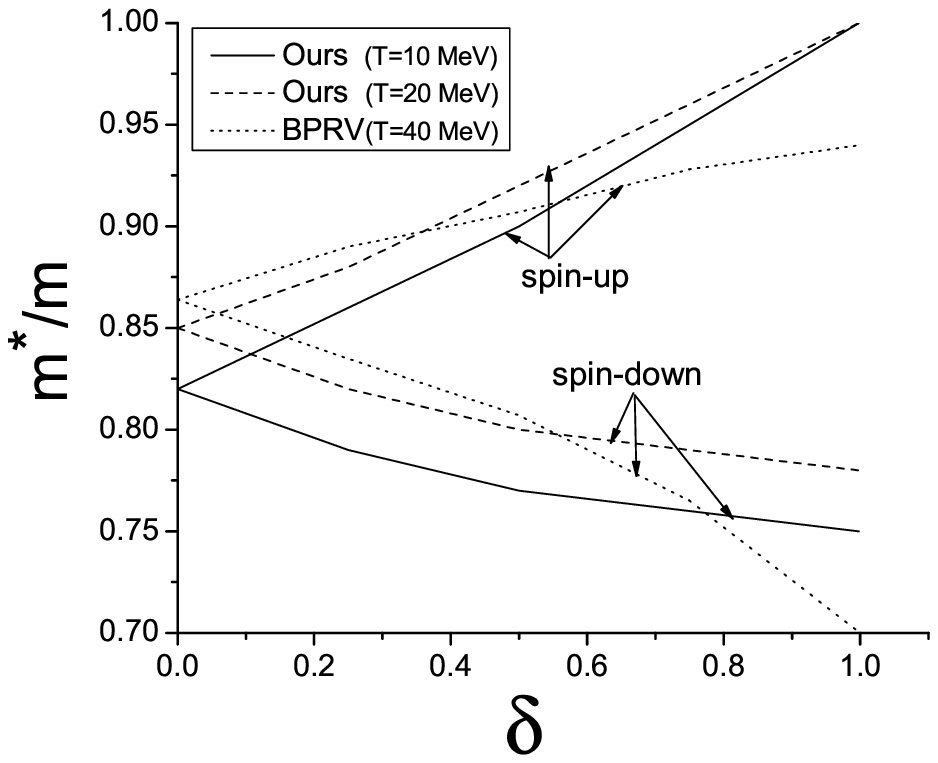}

 \caption{Our results for the effective mass of spin-up
 and spin-down
 neutrons versus spin polarization ($\delta$) for
 density $\rho=0.16 fm^{-3}$ at $T=10$ and $T=20$ MeV.
 The results of BPRV calculations [31] are also given
 for comparison.}
\label{correlate)}
\end{figure}
\newpage
\begin{figure}

\includegraphics{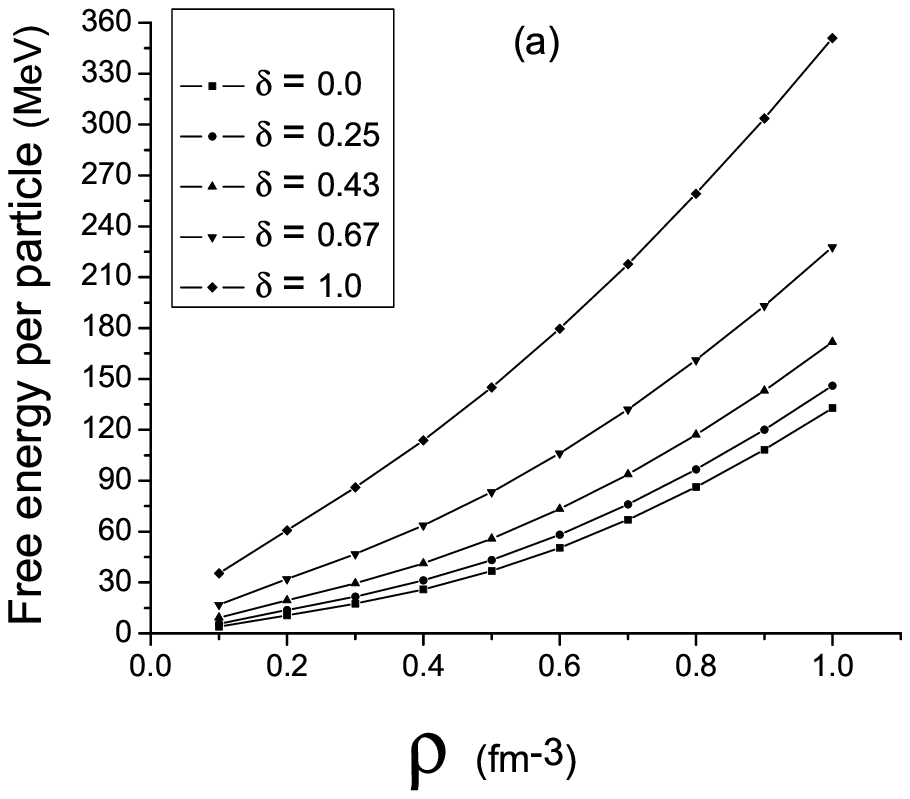}
\includegraphics{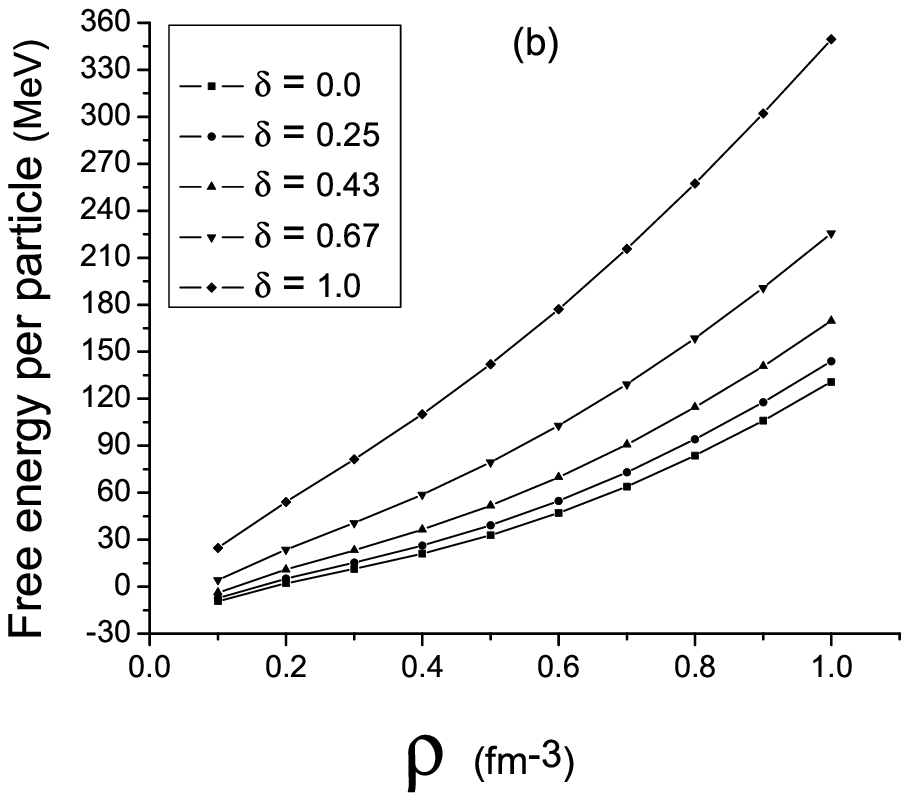}

 \caption{The free energy
per particle of polarized hot neutron matter as a function of the
total number density ($\rho$) for different values of the spin
polarization ($\delta$) at $T=10$ (a) and $T=20$ MeV (b).}
\label{correlate)}
\end{figure}

\newpage
\begin{figure}

\includegraphics{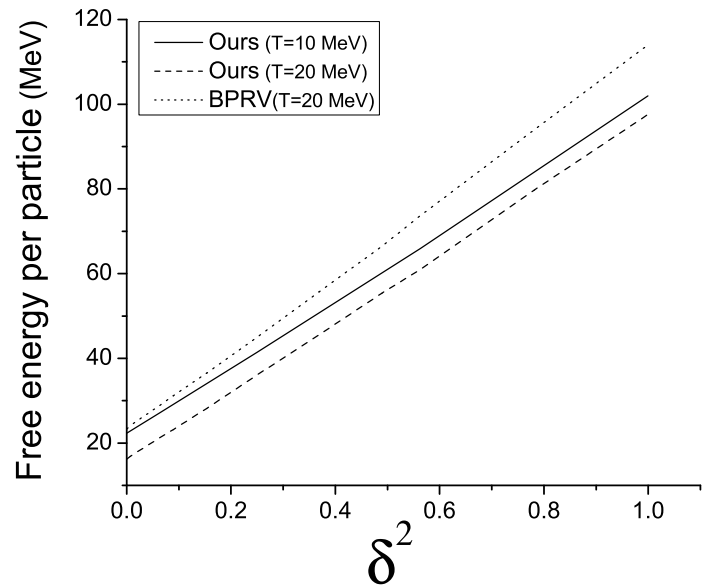}

 \caption{Our results for the free energy  as a
function of the quadratic spin polarization ($\delta^2$) at $T=10$
MeV (full curve) and $T=20$ MeV (dashed curve) for $\rho=0.36
fm^{-3}$. The results of BPRV [31] (dashed curve) are also given
for comparison.} \label{correlate)}
\end{figure}
\newpage
\begin{figure}

\includegraphics{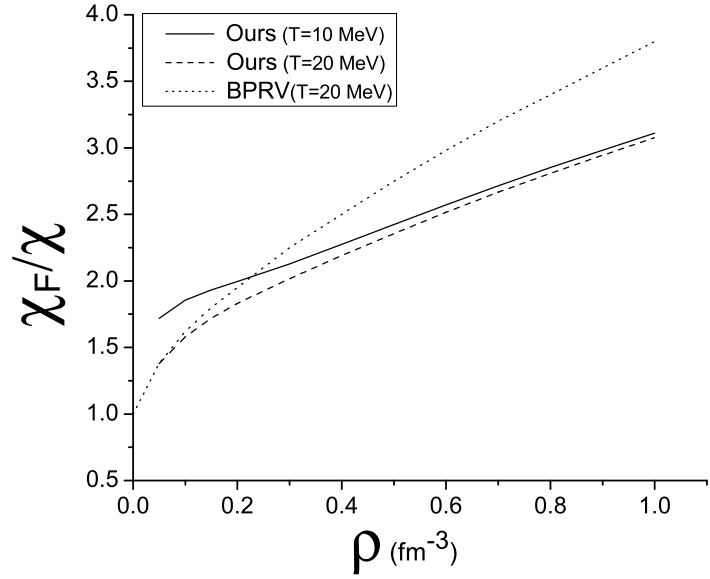}

 \caption{The magnetic
susceptibility of the hot neutron matter versus total number
density ($\rho$) at $T=10$ MeV (full curve) and $T=20$ MeV (dashed
curve). The results of BPRV [31] (dotted curve) are also given for
comparison .} \label{correlate)}
\end{figure}

\newpage
\begin{figure}

\includegraphics{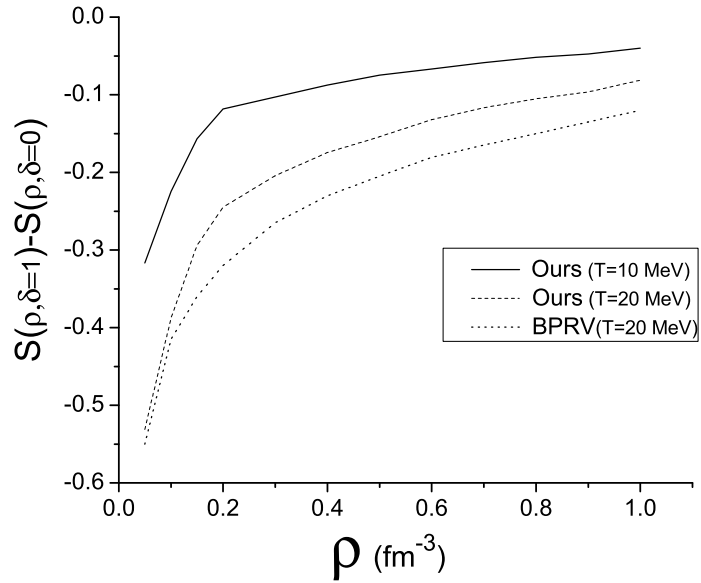}

 \caption{As Fig. 4 but for the entropy difference of
fully polarized and unpolarized cases  .} \label{correlate)}
\end{figure}
\newpage
\begin{figure}

\includegraphics{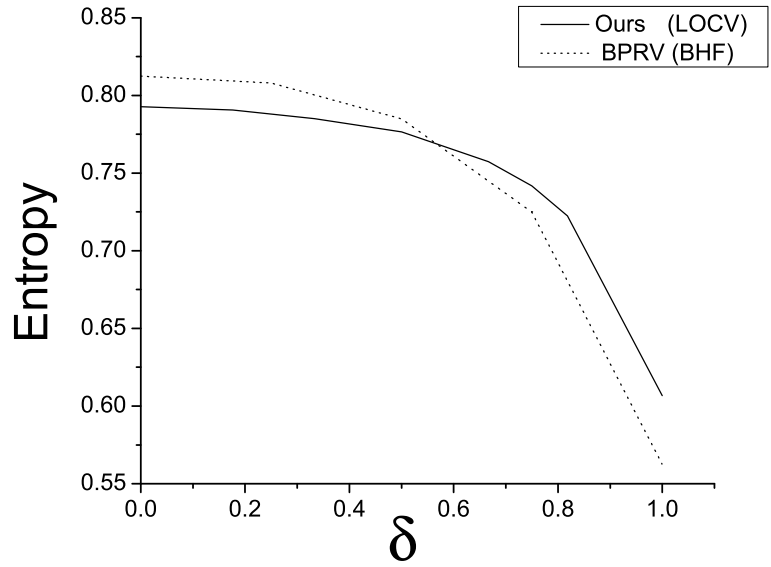}

 \caption{Our results (full curve) for the entropy per particle
                as a function of the spin polarization ($\delta$) at $T=20$
MeV and $\rho=0.32 fm^{-3}$. The results of BPRV [31] (dashed
curve) are also given for comparison.} \label{correlate)}
\end{figure}

\newpage
\begin{figure}

\includegraphics{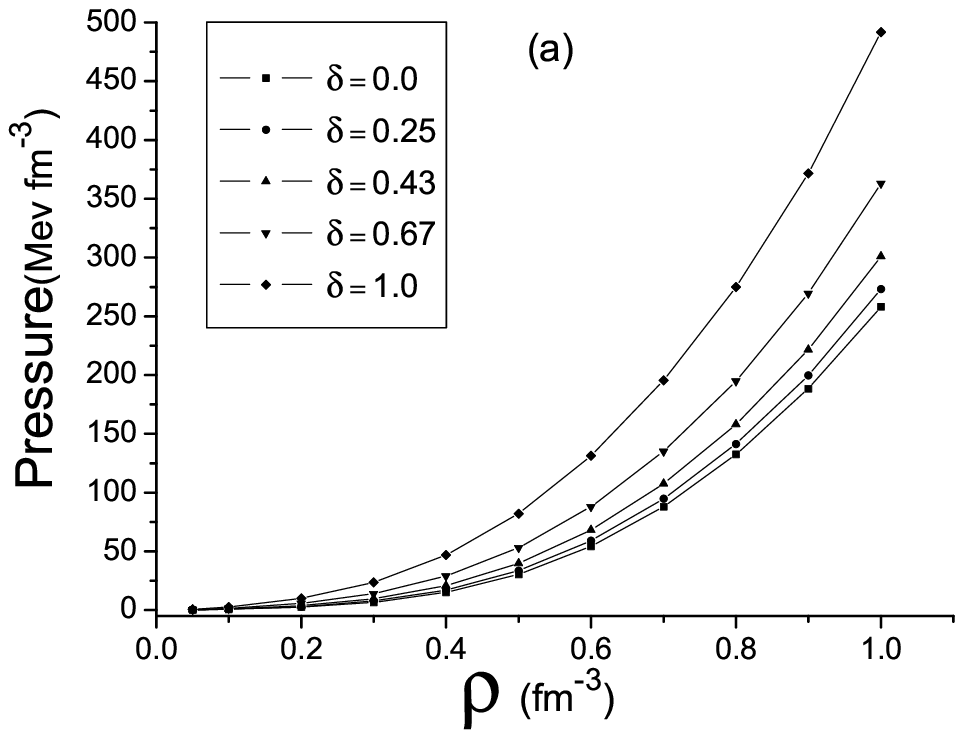}
\includegraphics{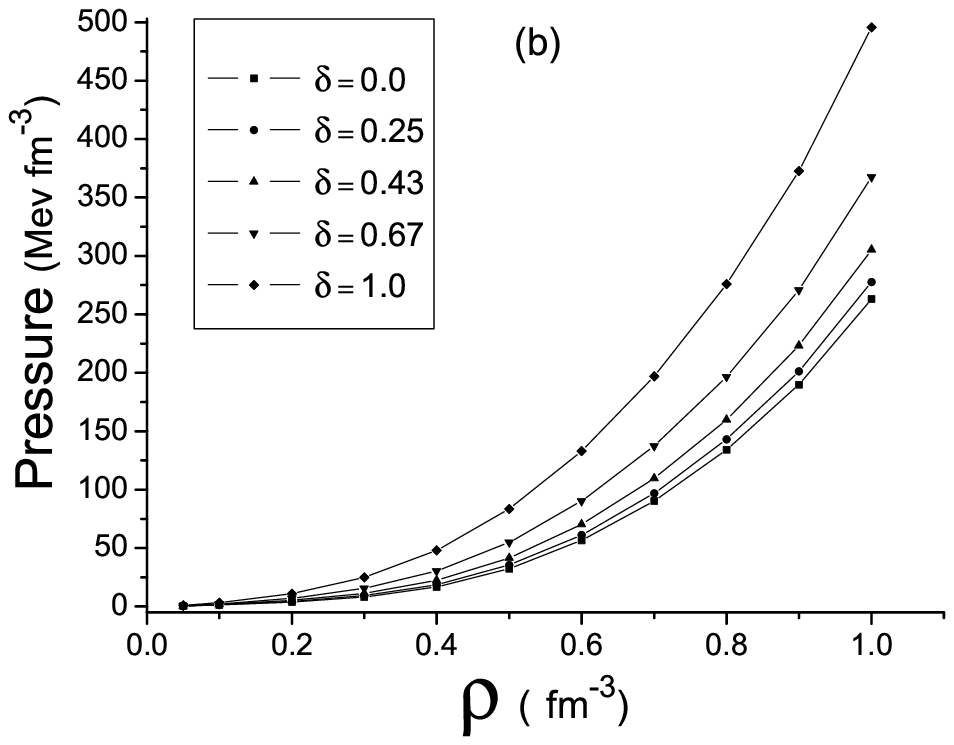}

 \caption{The equation of state of the hot neutron matter
              for different values of the spin polarization ($\delta$)
              at $T=10$ (a) and $T=20$ MeV (b).}
\label{correlate)}
\end{figure}


\end{document}